\documentclass[aps,showpacs,prx,amssymb,amsmath,twocolumn]{revtex4-2}
\usepackage{graphicx}
\usepackage{amssymb}
\usepackage{amsmath}
\usepackage{amsthm}
\usepackage{bm}
\usepackage{xcolor} 
\usepackage{array}
\usepackage{multirow}
\usepackage{tabularx}
\usepackage{booktabs}
\usepackage{textcomp}
\usepackage{mathtools}

\usepackage{bbm}
\usepackage{cancel}
\usepackage{comment}

\newcommand{\ket}[1]{\left|#1\right\rangle}

\theoremstyle{definition}

\begin{document}
\title{Rydberg Quantum Wires for  Maximum Independent Set Problems \\with Nonplanar and High-Degree Graphs}

\author{Minhyuk Kim, Kangheun Kim, Jaeyong Hwang, Eun-Gook Moon, and Jaewook Ahn}
\email{jwahn@kaist.ac.kr}
\address{Department of Physics, KAIST, Daejeon 34141, Republic of Korea}

\date{\today}

\begin{abstract} \noindent
One prominent application of near-term quantum computing devices is to solve combinatorial optimization such as non-deterministic polynomial-time hard (NP-hard) problems. Here we present experiments with Rydberg atoms to solve one of the NP-hard problems, the maximum independent set (MIS) of graphs. We introduce the Rydberg quantum wire scheme with auxiliary atoms to engineer long-ranged networks of qubit atoms. Three-dimensional (3D) Rydberg-atom arrays are constructed, overcoming the intrinsic limitations of two-dimensional arrays. We demonstrate Kuratowski subgraphs and a six-degree graph, which are the essentials of non-planar and high-degree graphs. Their MIS solutions are obtained by realizing a programmable quantum simulator with the quantum-wired 3D arrays. Our construction provides a way to engineer many-body entanglement, taking a step toward quantum advantages in combinatorial optimization.
\end{abstract}

\maketitle

\noindent
One of the latest efforts in quantum computation research is to use a
sizable quantum many-body system~\cite{Ebadi2021, Arute2019, Monroe2021, Scholl2020} to solve non-deterministic polynomial-time (NP)-optimization problems~\cite{Chen2011, Yarkoni2018, Centrone2021,Lucas2014}. Combinatorial optimization problems are to find an optimal solution from feasible solutions: for example, the maximum-independent-set (MIS) problem seeks to find an independent vertex set of maximal size for a graph~\cite{Korte2017}. While the MIS problem is classical by definition, their computational complexity (NP-complete) makes it intractable to classical Turing machines~\cite{Barahona1982,Arora2009}. Alternatively, if a quantum many-body system allows an intrinsic mapping of the problem to, for example, the many-body ground state, its evolution might be engineerable for the benefit of computational speedup~\cite{Farhi2001,Dickson2011}.

Rydberg atom systems provide an intrinsic Hamiltonian for the MIS problem~\cite{Lucas2014, Pichler2018}, of particular relevance in the context of the present paper. 
Consider $N$ atoms are arranged to a graph $G= G(V,E)$ with vertices ($V$) and edges ($E$), representing atoms and Rydberg-blockaded atom pairs, respectively. Their many-body ground state gives the MIS solution of $G$, $ \mathbb{M}(G)$~\cite{Bernien2017, Scholl2020}. The Hamiltonian is approximately given by
\begin{equation}
\hat{H}_{G} = U\sum_{(j,k)\in E}\hat n_{j}\hat n_{k} - \frac{\hbar\Delta}{2 } \sum_{j \in V} \hat \sigma_j^z \nonumber
\label{H}
\end{equation}
where $U$ is the nearest-neighbor interaction, $\Delta$ is the laser detuning, and $\hat n=(\hat \sigma^{z}+1)/2$ is the Rydberg excitation. The configuration $n=1$ ($n=0$) is for the Rydberg (ground) state of each atom, respectively~\cite{Labuhn2016}. It is easy to find that the many-body ground state of $\hat{H}_{G}$ gives $\mathbb{M}(G)$ with anti-ferromagnetic strong coupling  ($U>\hbar\Delta$) and positive detuning field ($\Delta>0$). Thus, quantum simulations for the ground state of the Schr\"{o}dinger equation, $\hat{H}_{G} | \mathbb{M}(G) \rangle = E_{g}(G) | \mathbb{M}(G) \rangle$, find the MIS solutions.  
\begin{figure}[thb]
  \centering
\includegraphics[width=0.45\textwidth]{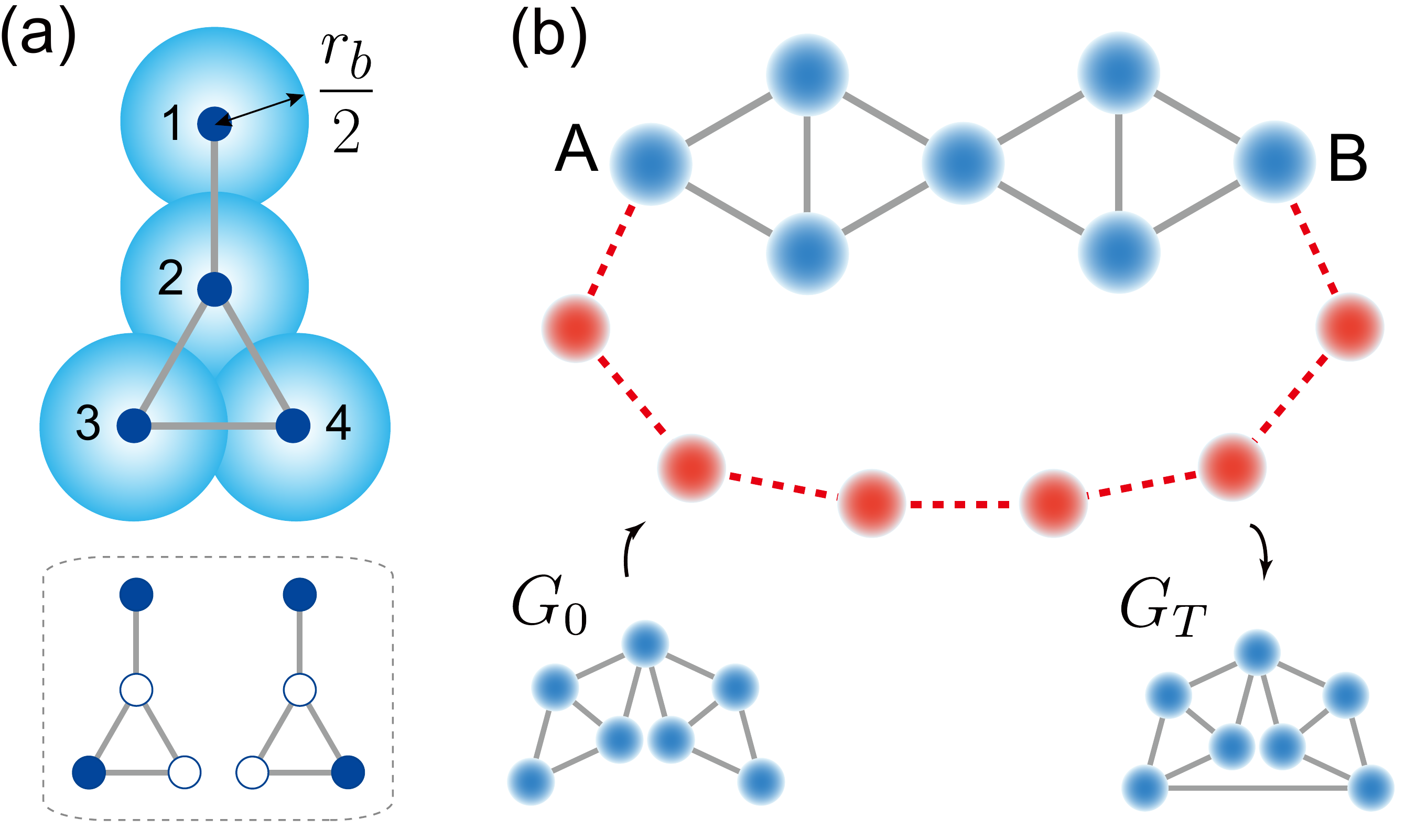}
\caption{(a) Graph representation of a Rydberg-atom array: Four ($N=4$) atoms are arranged in the nearest-neighbor Rydberg-blockade regime for $G(V,E)=$ the {\bf 3-pan} graph of $V=\{1,2,3,4\}$ and $E=\{\{1,2\},\{2,3\},\{2,4\},\{3,4\}\}$, of which the MIS solution is $\mathbb{M}(G)=\{\{1,3\},\{1,4\}\}$. (b) Rydberg quantum-wire concept: Qubit atoms (blue) are arranged to represent an initial graph $G_0={\bf X_{101}}$. When a chain of auxiliary atoms (red) makes a new edge between nonadjacent atoms (vertices A and B of $G_0$), the combined graph $G_{0+w}$ (qubit and wire atoms) constructs a target graph $G_T$, the {\bf Moser spindle} graph, by sharing the same MIS solutions.}
\label{fig1}
\end{figure}
In Fig.~\ref{fig1}(a), an example of two-dimensional (2D) Rydberg array simulations is presented with the 4-vertex graph, $G=$ the {\bf 3-pan} graph in the nomenclature of ISGCI (Information System on Graph Classes and their Inclusion)~\cite{ISGCI}. The atom arrangement, illustrated as the numbering in Fig.~\ref{fig1}(a), is suitable to utilize the Rydberg blockade with the radius, $r_b$, and its ground state $(\ket{1010}+\ket{1001})/\sqrt{2}$ gives the MIS solution, $\mathbb{M}(${\rm {\bf 3-pan}}$)=\{\{1,3\}, \{1,4\}\}$ by counting the $n=1$ atoms. 

We remark two intrinsic limitations of 2D Rydberg-atom arrays for the MIS problems. First, non-planar graphs cannot be simulated by 2D Rydberg atoms, as the mathematical theorem by Kuratowski indicates~\cite{Kuratowski1930}. Second, graphs with high-degree vertices cannot be encoded, because the size of Rydberg-atom interactions is set by the blockade radius. In this work, we demonstrate the two limitations are overcome by introducing a new scheme of quantum wires to Rydberg-atom arrays. 

Quantum wires are a chain of auxiliary wire atoms, proposed to mediate strong interactions between distant atoms in such a way that a complex target graph, $G_T$, can be synthesized from a simple initial graph, $G_0$. Such quantum wires are in principle realizable by using local addressing fields~\cite{Qiu2020}, but implementing them in the current Rydberg atom experiments is daunting. We, instead, propose an alternative quantum-wire scheme, termed as Rydberg quantum wire, which require no local addressing. Our scheme, as  to be shown below, demonstrates that the MIS problems with non-planar or high-degree graphs are readily accessible. 

The Rydberg quantum wire scheme consists of the three steps, construction of a wired array, quantum simulation for a MIS problem, and projection of the wire information. The construction step is illustrated in Fig.~\ref{fig1}(b), in which the quantum wire of an even number ($M=6$) of wire atoms (red spheres) is added to the initial graph, $G_0={\bf X_{101}}$ (blue spheres), to couple the $A$ and $B$ qubit atoms. If the quantum wire is treated as an edge, the combined graph $G_{0+w}$ is equivalent to the target graph, $G_T=$ the {\bf Moser spindle} graph. In the quantum simulation step, the ground state of the Hamiltonian is obtained by solving the Schr\"{o}dinger equation, $\hat{H}_{G_{0+w}} | \mathbb{M}(G_{0+w}) \rangle = E_{g}(G_{0+w}) | \mathbb{M}(G_{0+w}) \rangle$ experimentally, for example, with quantum annealing. Note that the size of the Hilbert space of $G_T$ is different from the one of $G_{0+w}$ by the  factor of $2^M$, where $M$ is the number of added atoms for the quantum wire. Thus, additional operations of quantum states of $G_{0+w}$ is performed in the final step. The MIS solution of a target graph, $\mathbb{M}(G_{T}) $, can be schematically written as, 
\begin{eqnarray}
| \mathbb{M}(G_{T}) \rangle = \mathbb{P}_F \mathbb{P}_H  \big[ | \mathbb{M}(G_{0+w}) \rangle\big]. \nonumber
\end{eqnarray}
The first operation ($\mathbb{P}_H$) is to make a projection of quantum states in the enlarged Hilbert space of $G_{0+w}$ onto the Hilbert space of the target graph $G_{T}$, which can be readily done by measuring the qubit information of $G_{T}$. Hereafter, we introduce the bar notation to specify the projection, for example, $\mathbb{M}(G) \rightarrow \mathbb{\bar{M}}(G)$. The second operation ($\mathbb{P}_F$) is to remove configurations with frustration between qubits at the boundaries, when some elements $\mathbb{F}(G_{0+w})$ of $\mathbb{\bar{M}}(G_{0+w})$ violate the Rydberg blockade condition. Then, the MIS solution of $G_T$ is obtained as
\begin{eqnarray}
 \mathbb{M}(G_T)  &=& \mathbb{\bar{M}}(G_{0+w}) - \mathbb{\bar{F}}(G_{0+w}) \label{eqn}
 \end{eqnarray}
(see Appendix for detailed discussions). It is noted that our Rydberg quantum wire scheme utilizes quantum entanglement of the two quantum many-body systems, original graph and wire. Namely, qubits of the two systems become entangled so that the MIS solution of a target graph is accessible. 

\begin{figure*}[htb]
\centering
\includegraphics[width=1\textwidth]{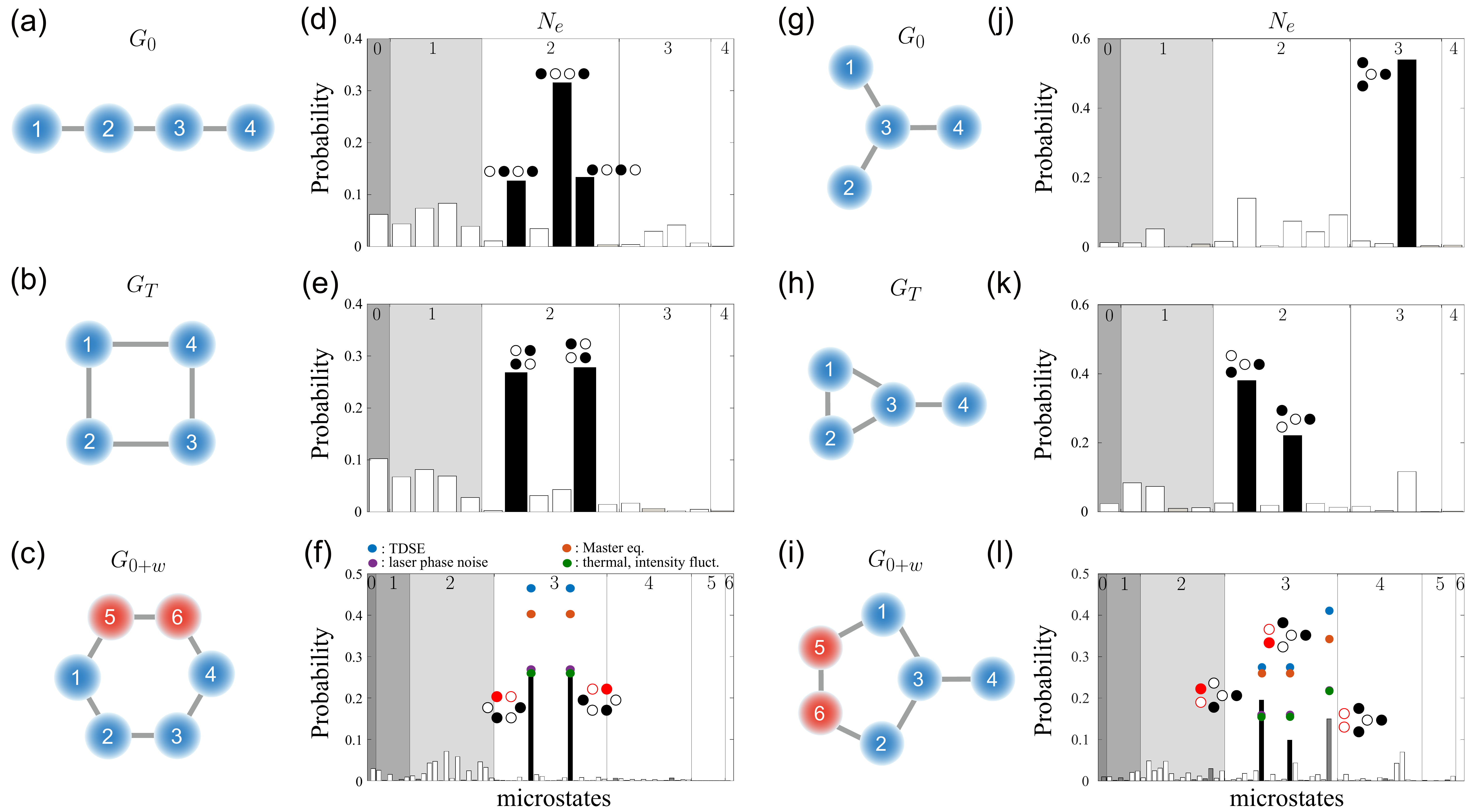}
\caption{Experimental tests of Rydberg quantum wires.  (a-c) A test graph set, $\{G_0, G_T, G_{0+w}\}$, for the algorithm with no frustration, $\mathbb{M}(G_T)=\mathbb{
\bar{M}}(G_{0+w})$: (a) $G_0={\bf P_4}$, the four-vertex path graph; (b) $G_T={\bf C_4}$, the four-vertex circle graph; (c) $G_{0+w}={\bf C_6}$, the six-vertex circle graph. (d-f) Experimental MIS solutions (filled bars) in the probabilities of all qubit-atom states in the bare-atom basis enumerated and sorted by $N_e=\sum_{j\in V} n_j$ (the total Rydberg excitation): (d) $\mathbb{M}(G_0)=\{\{2,4\},\{1,4\}, \{1,3\}\}$; (e) $\mathbb{M}(G_T)=\{\{2,4\},\{1,3\}\}$; and (f) $\mathbb{\bar{M}}(G_{0+w})=\{\{2,4\},\{1,3\}\}$. (g-l) A test graph with the frustration, $\mathbb{M}(G_T)=\mathbb{\bar{M}}(G_{0+w})- \mathbb{F}(G_{0+w})$: (g) $G_0={\bf S_4}$, the four-atom star graph; (h) $G_T=$ the {\bf 3-pan} graph; and (i) $G_{0+w}=$ the {\bf 5-pan} graph. (j-l) Experimental MIS solutions: (j) $\mathbb{M}(G_0)=\{\{1,2,4\}\}$; (k) $\mathbb{M}(G_T)=\{\{2,4\},\{1,4\}\}$; and (l) $\mathbb{\bar{M}}(G_{0+w})=\{\{2,4\},\{1,4\}, \{1,2,4\}\}$. In (d-f) and (j-l), state preparation and measurement errors of $P(0|1)=0.03$ and $P(1|0)=0.18$ are taken into account. In (f,l), numerical simulation for possible experimental errors are  estimated for comparison.}
\label{fig2}
\end{figure*}
 
Quantum simulation of MIS problems is performed with quantum annealing of 3D atom arrays~\cite{Song2021}. In experiments, neutral $^{87}$Rb atoms are arranged in free space in such a way that all nearest-neighbor atom pairs, which describe the edges of the graphs, are kept at a fixed interatomic distance $d$ smaller than the Rydberg blockade radius, i.e., $d<r_b=(C_6/\hbar\Omega)^{1/6}=9.8$~$\mu$m, and that all other atom pairs, which are not connected by edges, are at longer distances~\cite{Song2021,Kim2020}. The ground state $\ket{5S_{1/2}, F=2, m_F = 2} = \ket{n=0}$ and the Rydberg state $\ket{71S_{1/2}, m_J = 1/2} = \ket{n=1}$ of each atom are used for the qubit two-state system. An effective Hamiltonian of the 3D atom arrays is
\begin{equation}
\hat{H}(t) = U\sum_{(j,k)\in E}\hat n_{j}\hat n_{k} - \frac{\hbar}{2} \sum_{j \in V} \left( \delta(t) \hat \sigma^{z}_{j} -\Omega(t) \hat \sigma^x_{j}\right),
\label{Ht}
\end{equation}
where the Pauli matrices with the two states at sites $j,k$ are introduced with $\hat{n}_{j,k} = (\hat{\sigma}_{j,k}^z+1)/2$. Each term on the right hand side describes the van der Waals interaction at the fixed distance $d$,  the time-dependent detuning, and the time-dependent Rabi frequency, respectively.
Initially, the atoms are prepared in the paramagnetic down spins at $t=0$, $\ket{00\cdots0}$, with $\delta(0)=-\Delta_i<0$ and $\Omega(0)=0$. To find the MIS solutions of $G$, these atoms are quasi-adiabatically driven to the many-body ground state of $\hat H_G$, by turning on and off the Rabi frequency while the detuning is gradually increased to $\delta(t=t_f)=\Delta_f<U$~\cite{Song2021} (see Appendix for details).

We first consider in Fig.~\ref{fig2} experimental tests of Rydberg quantum wires for the cases with and without the frustration, ${\mathbb{\bar F}(G_{0+w})}=\emptyset$ and $\neq\emptyset$ in Eq.~\eqref{eqn}, respectively. For graphs without the frustration, we consider the initial graph, $G_0={\bf P}_4$ in Fig.~\ref{fig2}(a), and the target graph, $G_T={\bf C}_4$ in Fig.~\ref{fig2}(b). The construction of the wired graph, $G_{0+w}={\bf C}_6$, is done by adding an $M=2$ Rydberg quantum wire (red spheres) as shown in Fig.~\ref{fig2}(c). Quantum simulations observe high-population states as in Figs.~\ref{fig2}(d,e,f), of which the MIS solutions are summarized as $\mathbb{M}(G_0)=\{\{2,4\}, \{1,4\},\{1,3\}\}$, $\mathbb{\bar{M}}(G_{0+w})=\{\{2,4\}, \{1,3\}\}$, and
$\mathbb{M}(G_{T})=\{\{2,4\}, \{1,3\}\}$.
It is easy to verify that $\mathbb{F}(G_{0+w})= \emptyset$ and $\mathbb{M}(G_T)  = \mathbb{\bar{M}}(G_{0+w})$, satisfying the MIS solution in Eq.~\eqref{eqn}. Note that the population difference among the MIS solutions of $G_0$, in Fig.~\ref{fig2}(d), is due to the fact that the quantum annealing results in a coherent superposition of the MIS solutions, $\ket{ \mathbb{M}(G_{0}) \rangle}=(\ket{0101}+\ket{1010})/\sqrt{2}+\ket{0110}/2$.

For graphs with the frustration, i.e., $\mathbb{F}(G_{0+w}) \neq \emptyset$, we consider the initial and target graphs, $G_0={\bf S_4}$ and $G_T=$ {\bf 3-pan} as in Fig.~\ref{fig2}(g,h). The wired graph is $G_{0+w}=$ {\bf 5-pan}, constructed with the Rydberg quantum wire as shown in Fig.~\ref{fig2}(i). Quantum simulations of $G_0$, $G_T$, and $G_{0+w}$ are respectively shown in Figs.~\ref{fig2}(j,k,l), of which the high-population states are given by
$\mathbb{M}(G_0)=\{\{1,2,4\}\}$, $\mathbb{M}(G_T)=\{\{1,4\},\{2,4\}\}$, and
$\mathbb{\bar{M}}(G_{0+w})=\{\{1,4\},\{2,4\}, \{1,2,4\}\}$. 
The results also confirm the MIS solution, $\mathbb{M}(G_T)  = \mathbb{\bar{M}}(G_{0+w})-\mathbb{\bar{F}}(G_{0+w})$, where $\mathbb{\bar{F}}(G_{0+w})= \{{\{1,2,4\}}\}$ is the frustrated configuration.
 
\begin{figure*}[htb]
\centering
\includegraphics[width=0.95\textwidth]{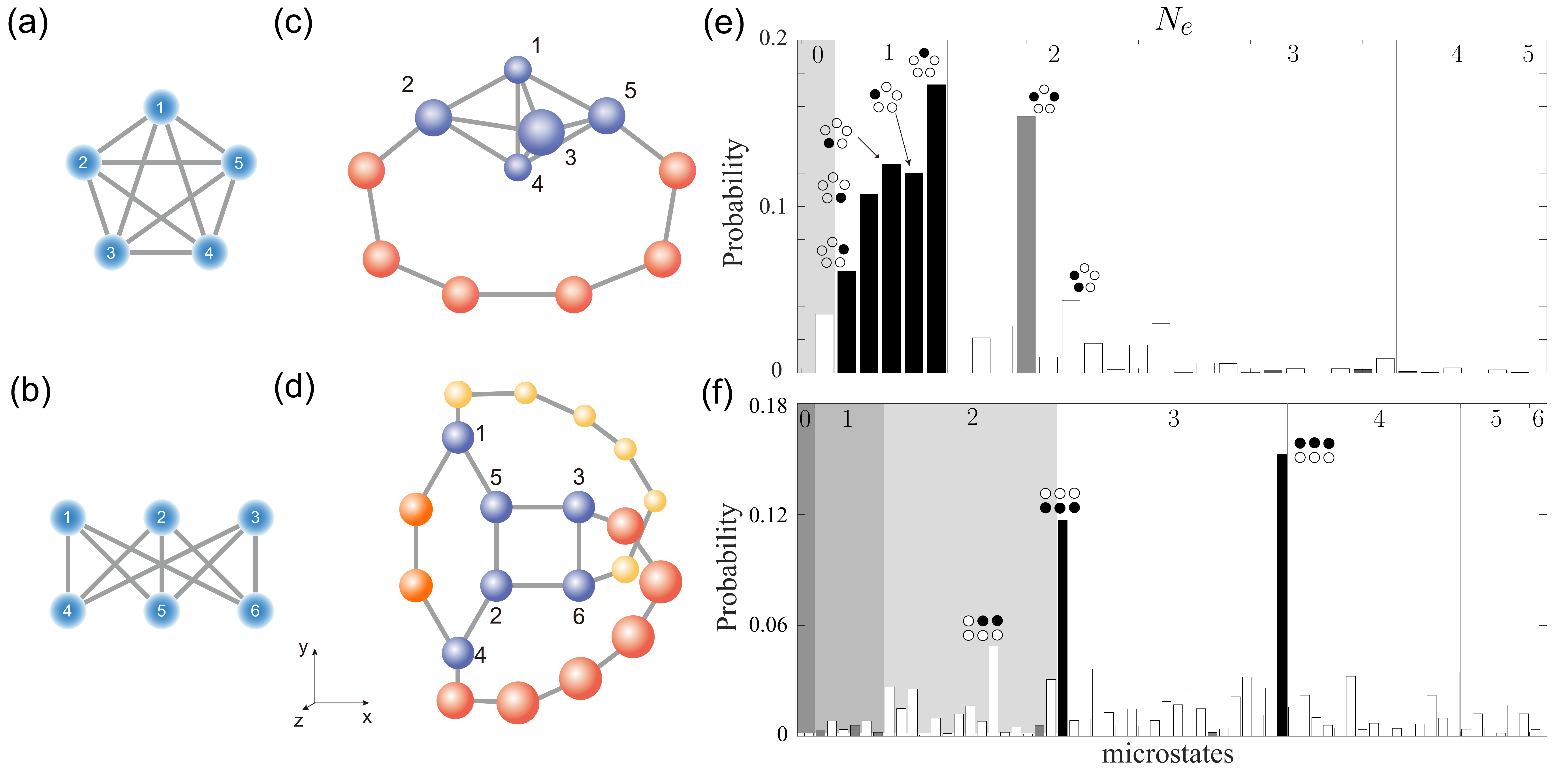}
\caption{Nonplanar graphs. Experimental construction of Kuratowski subgraphs: (a) ${\bf K_5}$. (b) ${\bf K_{3,3}}$. (c) Rydberg quantum-wire implementation of ${\bf K_5^{exp}}$ with ${\bf K_5}\mbox{-}e$ (blue) and a ${\bf P_6}$ quantum wire (red). (d) ${\bf K_{3,3}^{exp}}$ with the {\bf A} graph (blue) and three quantum wires (yellow, orange, and red). (e) Experimental MIS solutions $\mathbb{\bar M}({\bf K_5^{exp}})=\{\{1\},\{2\},\cdots, \{5\}, \{2,5\}\}$, in which \{2,5\} has a frustrated quantum wire. (f) $\mathbb{\bar M}({\bf K_{3,3}^{exp}})=\{\{1,2,3\},\{4,5,6\}\}$.}
\label{fig3}
\end{figure*}

Next, we consider nonplanar graphs, focusing on Kuratowski subgraphs, ${\bf K_5}$ and ${\bf K_{3,3}}$. The seminal work by Kuratowski shows that a graph $G$ is nonplanar if and only if $G$ contains any of these two Kuratowski subgraphs~\cite{Kuratowski1930}.  ${\bf K_5}$ in Fig.~\ref{fig3}(a) is a complete graph with each vertex edged to all other vertices and ${\bf K_{3,3}}$ in Fig.~\ref{fig3}(b) is a bipartite graph with three vertices on one side completely connected to the vertices on the other side. It has been shown that both of these graphs require quantum wiring as well as 3D atom arrangements~\cite{Barredo2018,Lee2016}. 

We construct wired graphs, $G_{0+w}={\bf K_5^{exp}}$ and ${\bf K_{3,3}^{exp}}$, as shown in Figs.~\ref{fig3}(c,d), to simulate the target graphs $G_T= {\bf K_5}$ and ${\bf K_{3,3}}$, respectively. In Fig.~\ref{fig3}(c), the initial graph is $G_0={\bf K_5}\mbox{-}e$ of five qubit atoms (blue spheres) in the tetrahedral configuration, of which the two qubit atoms (2 and 5) are coupled by the quantum wire of six wire atoms (red spheres). With the constructed ${\bf K_5^{exp}}$, quantum annealing is performed and the result is shown in Fig.~\ref{fig3}(e). Six peaks are observed corresponding to the five singly-excited solutions (black bars), $\{1\}$, $\{2\}$, $\cdots$, $\{5\}$, and a doubly-excited $\{2,5\}$ (gray bar). The last solution, $\{2,5\}$, is frustrated, and the MIS solution of ${\bf K_5}$ is obtained by Eq.~\eqref{eqn}, $\mathbb{M}({\bf K_5})=\{ \{1\},\{2\}, \cdots \{5\}\}$.
We note that the uneven probabilities of the singly-excited states are attributed to the difference of their higher-order interactions, which are ignored in Eq.~\eqref{H}, to other qubit atoms and the quantum wire, which are mainly due to the asymmetries of physical implementations (for example, atom 3 is kept closer to atom 5 than 2 to avoid the interference of holographic optical potentials~\cite{Barredo2018}). 
In Fig.~\ref{fig3}(d) for ${\bf K_{3,3}^{exp}}$, the initial graph is the {\bf A} graph (blue) and three quantum wires (yellow, orange, and red) are used to connect  $(1,6)$, $(1,4)$, and $(3,4)$ pairs. Quantum annealing of the constructed ${\bf K_{3,3}^{exp}}$ results in two peaks as shown in Fig.~\ref{fig3}(f), corresponding to the two MIS solutions  (black bars), $\{1,2,3\}$ and $\{4,5,6\}$ of ${\bf K_{3,3}^{exp}}$. Both the solutions are not frustrated and the Rydberg quantum-wire algorithm in Eq.~\eqref{eqn} experimentally finds the MIS solution of ${\bf K_{3,3}}$ as $\mathbb{M}({\bf K_{3,3}}) =\{ \{1,2,3\}, \{4,5,6\} \}$.

\begin{figure}[htb]
\centering
\includegraphics[width=0.48\textwidth]{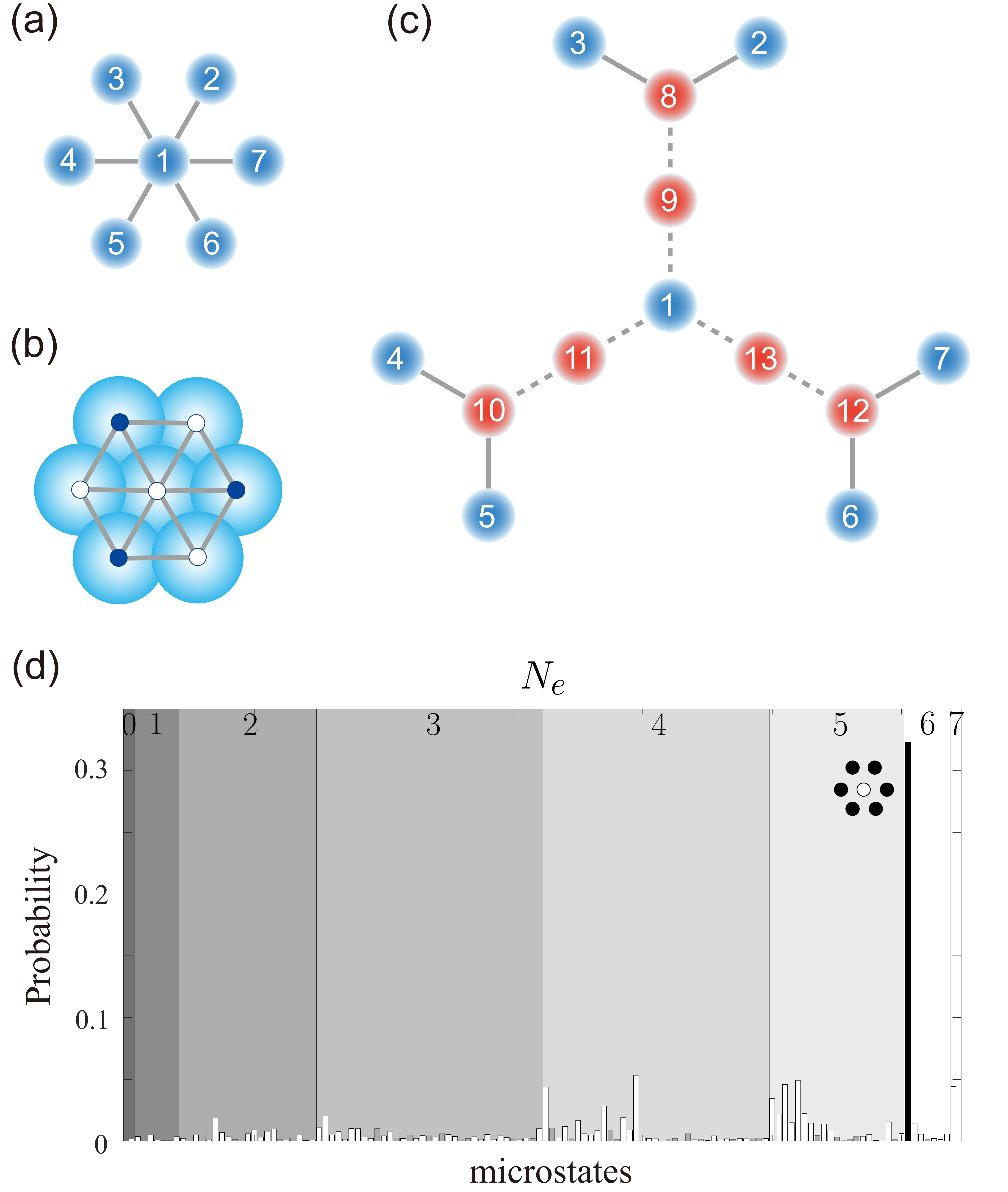}
\caption{Vertex-splitting demonstration: (a) A star graph ${\bf S_6}$ of the six-degree center vertex. (b) 2D implementation of ${\bf S_6}$ results in a wheel graph ${\bf W_7}$.
(c) Three quantum wires are used to split the center 6-degree vertex to three extended 3-degree vertices. (d) Experimental MIS solution $\mathbb{\bar{M}}({\bf S_6^{exp}})=\{\{2,3,\cdots,7\}\}=\mathbb{M}({\bf S_6})$.}
\label{fig4}
\end{figure}

As an additional application of Rydberg quantum wires, we consider a graph of high-degree vertex in Fig.~\ref{fig4}. Implementation of a high-degree vertex is crucial in quantum simulations with Rydberg-atom arrays. For example, the star graph ${\bf S_6}$ in Fig.~\ref{fig4}(a), which has a six-degree vertex at the center, cannot be simulated in any 2D arrays without a quantum wire scheme, because a naive 2D implementation of atoms results in a different graph, the wheel graph ${\bf W_7}$, as shown in Fig.~\ref{fig4}(b). Our strategy is to reduce the degree of the high-degree vertex, by using the Rydberg quantum wires, also known as vertex-splitting in graph theory~\cite{Garey1977}. 
As shown in  Fig.~\ref{fig4}(c), the 13-vertex extended tree-like graph (red and blue spheres), ${\bf S_6^{exp}}$, is constructed from the initial graph, $7{\bf K_1}$ of seven isolated vertices. Three quantum wires are used to split the 6-degree center vertex by adding the three 3-degree vertices. Quantum annealing results of the as-constructed ${\bf S_6^{exp}}$ are shown in Fig.~\ref{fig4}(d). A single peak is observed corresponding to the MIS solution of ${\bf S_6^{exp}}$, which is not frustrated {($\mathbb{\bar{F}}(G_{0+w})=\emptyset$)}, i.e., $\mathbb{\bar M}({\bf S_6^{exp}}) = \{ \{2,3,\cdots,7\} \}$, and the Rydberg quantum wire scheme successfully constructs the high-degree graph.

In summary, we have proposed and experimentally demonstrated the Rydberg quantum wire scheme, which utilizes Rydberg many-body interactions along a chain of neutral atoms to program the complex connections of nonplanar and high-degree graphs necessary for general MIS problems. We have used 3D arrays of qubit and quantum-wire atoms, to construct the Kuratowski subgraphs, ${\bf K_5}$ and ${\bf K_{3,3}}$, and the six-degree graph, ${\bf S_6}$, and probed their many-body ground states using the near-adiabatic quantum annealing procedure. The observed ground-states of the quantum-wired systems have exhibited excellent agreements with, or algorithmically retrieved, the MIS solutions of the target graphs. Our demonstration suggests that a general graph of $N$-by-$N$ couplings is in principle implementable by Rydberg quantum wires, while  there remain unresolved issues such as limited physical resources~\cite{Preskill2018}, efficient many-body ground-state probing~\cite{Denchev2016}, and technical issues involved with tangled 3D wires. It is hoped that our quantum-wire scheme demonstrated for MIS problems shall be useful to further developments for other optimization problems. 

\begin{appendix}

\section{Quantum Mechanics of the wired MIS problems} \noindent
Our goal is to find the MIS solution of a target graph, $G_{T}$, which can be obtained by finding the ground state of the Schr\"{o}dinger equation of the target Hamiltonian, 
\begin{eqnarray}
&&\hat{H}_{T}  | \mathbb{M}(G_{T}) \rangle = E_{g}(G_{T}) | \mathbb{M}(G_{T}) \rangle. \nonumber 
\end{eqnarray}
However, the construction of $G_T$ is fundamentally limited by non-planar graphs and high-degree vertices. 

Our strategy to overcome the limitations is to investigate a wired graph, $G_{0+w}$, and obtain the MIS solution of $G_T$.
The basic fact is the relation, $\mathbb{M}(G_{T}) \subset \mathbb{\bar{M}}(G_{0+w})$, which can be easily shown by replacing an edge of $G_T$ by a quantum wire. 
The MIS solution of $G_{0+w}$ is obtained by solving the Schr\"{o}dinger equation, 
\begin{eqnarray}
&&\hat{H}_{0+w}  | \mathbb{M}(G_{0+w}) \rangle = E_{g}(G_{0+w}) | \mathbb{M}(G_{0+w}) \rangle. \nonumber 
\end{eqnarray}
The ground state can be written as 
\begin{eqnarray}
 | \mathbb{M}(G_{0+w}) \rangle  = \sum_j C_j | \Psi_j \rangle= \sum_j C_j | \psi_j \rangle \otimes | \phi_j \rangle. \nonumber
\end{eqnarray}
Note that  the qubit state $ | \psi_j \rangle$ (the combined state $| \Psi_j \rangle$) is located in the Hilbert space whose size is $2^{N }$ ($2^{N+M}$), where $N$ and $M$ are the numbers of qubits of the target graph and wires, respectively. 
The MIS solution of the wired graph is 
 \begin{eqnarray}
 \mathbb{M}(G_{0+w}) &=& \{  (\psi_1, \phi_1), (\psi_2, \psi_2), \cdots  \}. \nonumber 
 \end{eqnarray}

We introduce the projection operator, $\mathbb{P}_H$, defined as 
\begin{eqnarray}
\mathbb{P}_H | \Psi_j  \rangle = | \psi_j \rangle. \nonumber
\end{eqnarray}
Mathematically, a Fock space with different qubit numbers is necessary, and yet it is safe to use the above notation for the MIS problems. 
The projection state is
\begin{eqnarray}
 \mathbb{P}_H  | \mathbb{M}(G_{0+w}) \rangle  = \sum_j C_j | \psi_j \rangle  \nonumber
\end{eqnarray}
and the corresponding solution set is 
 \begin{eqnarray}
  \mathbb{\bar{M}}(G_{0+w}) &=& \{  \psi_1, \psi_2, \cdots, \tilde{\psi}_1, \tilde{\psi}_2, \cdots \}. \nonumber
\end{eqnarray}
The bar notation is introduced to specify  the projection. 
Some of the elements of $\mathbb{\bar{M}}(G_{0+w})$ are not a solution of the MIS problem of $G_T$, which are specified by the tilde notation. One can introduce the frustration function defined as $f(n_A, n_B) =n_A n_B$, where $n_A$ and $n_B$ are the number operator eigenvalues of the boundaries where a wire is connected. 
For an element of $\mathbb{\bar{M}}(G_{0+w})$,  the Rydberg blockade condition, $f(n_A, n_B)=1$, can be tested, and a set with the condition is obtained, $\mathbb{\bar{F}}(G_{0+w}) = \{ \tilde{\psi}_1, \tilde{\psi}_2, \cdots\}$. 
Then, one can introduce the second projection operator, $\mathbb{P}_F = \mathbb{I} - \sum_{a} | \tilde{\psi}_a \rangle \langle \tilde{\psi}_a |$, and the final state  is
\begin{eqnarray}
| \mathbb{M}(G_{T}) \rangle &=& \mathbb{P}_F \mathbb{P}_H  \big[ | \mathbb{M}(G_{0+w}) \rangle\big]. \nonumber 
\end{eqnarray}
Then, the solution set can be written as 
 \begin{eqnarray}
 \mathbb{M}(G_{T}) &=&\mathbb{\bar{M}}(G_{0+w}) -  \mathbb{\bar{F}}(G_{0+w})= \{  \psi_1 \psi_2, \cdots  \}. \label{eqA1}
 \end{eqnarray}
 
\section{Rydberg-atom quantum simulator} \noindent
Experiments of Rydberg quantum wires are performed with a 3D Rydberg-atom quantum simulator, which consists of a magneto-optical trap (MOT) of $^{87}$Rb atoms, an optical system for holographical optical tweezers, a laser system for Rydberg-atom excitations, and a single-atom detection system. 

The atoms are cooled in the MOT down to 30~$\mu$K by Doppler and polarization gradient cooling, and optically pumped to the ground hyperfine state $\ket{0}=\ket{5S_{1/2}, F=2, m_F=2}$. After the MOT is turned off, optical tweezers (far-off resonant optical dipole traps) are turned on to capture single atoms. Up to 250 optical tweezers are created at predetermined 3D locations of 5-10~$\mu$m spacing, by using an 820-nm laser (Ti:Sapphire CW laser of Avesta), a spatial light modulator (SLM, ODPDM512 of Meadowlark optics), and a microscope objective lens (Mitutoyo G Plan Apo 50$\times$). Each optical tweezer has the trap depth of 1~mK, the diameter of 2 $\mu$m, and the lifetime of 40(10)~s. 

To make a defect-free arrangement of $N$ atoms, we use $N$ optical tweezers on target atom positions and another $N$ optical tweezers as a reservoir around the targets. The atom occupations of the optical tweezers are determined by fluorescence imaging of the $\ket{5S_{1/2}, F=2}-\ket{5P_{3/2}, F=3}$ transition with an electron-multiplied-CCD camera and an electrically-tunable-lens (ETL, EL-16-40-TC of Optotune). The lateral and axial resolutions are 0.3~$\mu$m and 0.5~$\mu$m, respectively. After the occupations are checked, the captured atoms are rearranged by optical tweezers steered along a set of paths obtained by Hungarian algorithm~\cite{Lee2017}. 3D Gerchberg-Saxton  (GS) algorithm is used to program the dynamic holograms, of which the real-time calculation is performed  with a GPU (NVIDIA, Titan-X pascal). The 3D GS algorithm uses for fast convergence the weights of target site intensity feedback, and the initial phase pattern of SLM for every iteration utilizes the calculated phase pattern from the former iteration~\cite{Kim2019}. As example, a 35 times iteration takes about 700~ms and it is sufficient to move traps about 20~$\mu$m with an over 90 \% occupation probability of each site. The 3D atom positions of all experimental graphs are listed  in Table~\ref{tableS1}. 

\begin{table*}[tbh]
\caption{Atom positions of the graphs demonstrated in the main text. }
\centering
\begin{ruledtabular}
\begin{tabular}{c c l l l l}
Figure & Graph & Positions ($x, y, z$) ($\mu$m) & & &\\
\hline\hline
2a & ${\bf P_4}$  & 1 : (-7.0, 3.0, 0.0) & 2 : (-3.5, -3.0, 0.0) & 3 : (3.5, -3.0, 0.0) & 4 : (7.0, 3.0, 0.0) \\
\hline
2b& ${\bf C_4}$   & 1 : (-3.5, 3.5, 0.0) & 2 : (-3.5, -3.5, 0.0) & 3 : (3.5, -3.5, 0.0) & 4 : (3.5, 3.5, 0.0) \\
\hline
2c& ${\bf C_6}$  &  1 : (-7.0, 0.0, 0.0) & 2 : (-3.5, -6.1, 0.0) & 3 : (3.5, -6.1, 0.0)  & 4 : (7.0, 0.0, 0.0) \\  & & 5 : (-3.5, 6.1, 0.0) & 6 : (3.5, 6.1, 0.0) \\
\hline
2g& ${\bf S_4}$ & 1 : (-3.8, 6.5, 0.0) & 2 : (-3.8, -6.5, 0.0) & 3 : (0.0, 0.0, 0.0) &
 4 : (7.5, 0.0, 0.0) \\
\hline
2h& {\bf 3-pan} & 1 : (-6.5, 3.8, 0.0) & 2 : (-6.5, -3.8, 0.0) & 3 : (0.0, 0.0, 0.0) 
& 4 : (7.5, 0.0, 0.0) \\
\hline
2i& {\bf 5-pan} &  1 : (-3.8, 6.5, 0.0) & 2 : (-3.8, -6.5, 0.0) & 3 : (0.0, 0.0, 0.0) &
 4 : (7.5, 0.0, 0.0) \\ & & 5 : (-10.7, 3.8, 0.0) & 6 : (-10.7, -3.8, 0.0) \\
\hline
3a& ${\bf K_5}$ &  1 : (-2.0, 0.0, -2.1) & 2 : (-5.0, 0.0, 2.5) & 3 : (5.0, 0.0, -2.5) 
& 4 : (1.0, 2.9, 1.5) \\ 
& & 5 : (1.0, -3.5, 1.5) & 6 : (-9.6, 5.0, 4.0) & 7 : (-8.8, 12.1, 3.6) & 8 : (-5.1, 12.1, -0.8) \\
& & 9 : (1.0, 12.1, -3.8) & 10 : (8.1, 12.1, -4.7) & 11 : (8.8, 5.0, -5.1)  \\
\hline
3b& ${\bf K_{3,3}}$ &  1 : (-6.9, 8.0, 0.0) & 2 : (0.0, -4.0, 0.0) & 3 : (8.0, 4.0, 0.0) & 4 : (-6.9, -8.0, 0.0) \\
& & 5 : (0.0, 4.0, 0.0) & 6 : (8.0, -4.0, 0.0) & 7 : (-13.9, 4.0, 0.0) & 8 : (-13.9, -4.0, 0.0) \\
& & 9 : (-6.9, -13.7, -5.7) & 10 : (0.0, -16.0, -7.0) & 11 : (7.0, -13.0, -7.0) & 12 : (14.5, -10.5, -7.0) \\
& & 13 : (17.0, -3.0, -7.0) & 14 : (13.7, 4.0, -5.7) & 15 : (-6.9, 13.7, 5.7) & 16 : (0.0, 16.0, 7.0) \\ 
& & 17 : (7.0, 13.0, 7.0) & 18 : (14.5, 10.5, 7.0) & 19 : (17.0, 3.0, 7.0) & 20 : (13.7, -4.0, 5.7) \\
\hline
4c& ${\bf S_{6}}$ &  1 : (0.0, 0.0, 0.0) & 2 : (6.1, 17.5, 0.0) & 3 : (-6.1, 17.5, 0.0) & 4 : (-18.2, -3.5, 0.0) \\ 
& & 5 : (-12.1, -14.0, 0.0) & 6 : (12.1, -14.0, 0.0) & 7 : (18.2, -3.5, 0.0) & 8 : (0.0, 14.0, 0.0)\\
& & 9 : (-12.1, -7.0, 0.0) & 10 : (12.1, -7.0, 0.0) & 11 : (0.0, 7.0, 0.0) & 12 : (-6.1, -3.5, 0.0) \\
& & 13 : (6.1,-3.5,0.0) &  &  \\
\end{tabular}
\end{ruledtabular}
\label{tableS1}
\end{table*}

The atoms are two-photon excited from the ground state $\ket{0}=\ket{5S_{1/2}, F=2, m_F=2}$ to the Rydberg state $\ket{1}=\ket{71S_{1/2}, m_j=1/2}$ via the intermediate state $\ket{m}=\ket{5P_{3/2}, F=3, m_F=3}$. We use 780~nm (home-made extra-cavity diode laser) and 480~nm (TA-SHG Pro from Toptica) lasers for the $\ket{0}\rightarrow \ket{m}$ and $\ket{m}\rightarrow \ket{1}$ transitions, respectively. Both the lasers are frequency-stabilized to an ultra-low-expansion cavity (Stable laser systems, finesse 15000) by Pound-Drever-Hall method. The effective Rabi frequency of the two-photon transition is 
$\Omega=\Omega_{\rm 780}\Omega_{\rm 480}/2\Delta_m = 2\pi\times0.88(1)$~MHz, where $\Omega_{\rm 780}=2\pi\times50$~MHz and $\Omega_{\rm 480}=2\pi\times20$~MHz are the Rabi frequencies are the 780~nm and 480~nm lasers, respectively, and $\Delta_m = 2\pi\times570$~MHz is the intermediate detuning. The laser power of 780(480)~nm beam is 50~$\mu$W (550~mW) and the $1/e^2$ beam diameter is 180~$\mu$m (100~$\mu$m), which is sufficient to cover the atomic array we used with near-uniform Rabi frequencies less than 10~$\%$ deviation. The typical decay time of the single-atom Rabi oscillation is 7(2)~$\mu$s~\cite{Kim2020}.

For quantum annealing, the control parameters $\delta(t)$ and $\Omega(t)$ of the Hamiltonian $\hat{H}(t)$ in Eq.~\eqref{Ht} are adjusted from the paramagnetic phase condition, $\delta(t=0)=\Delta_i<0$ and $\Omega(t=0)=0$, to the MIS phase condition, $\delta(t=t_f)=\Delta_f<U$ and $\Omega(t=t_f)=0$. Within the time of $t_f = 4~\mu$s, we use three regions of operations. At the first region during $t : 0 \rightarrow t_f/10$, $\delta(t)$ is fixed to $\Delta_i$, and $\Omega(t)$ is linearly increased from zero to $\Omega_0$, for the initial state to evolve to the ground state of quantum Ising Hamiltonian. Then at the second region $\delta(t)$ is linearly increased from $\Delta_i<0$ to $\Delta_f>0$, while $\Omega(t)=\Omega_0$ is constant during $t : t_f/10 \rightarrow 9t_f/10$. Finally at the third region, during $t : 9t_f10 \rightarrow t_f$, $\Omega(t)$ gradually decreases to zero while $\delta(t)=\Delta_f$ is kept constant~\cite{Song2021}. Table~\ref{tableS2} shows the interatomic distance, $d$, initial detuning, $\Delta_i$, final detuning, $\Delta_f$, and measurement repetitions of each experiment. 

\begin{table}[h]
\caption{Experimental parameters}
\centering
\begin{ruledtabular}
\begin{tabular}{c c c c c c}
Figure & Graph & $d$~($\mu$m) & $\Delta_i$~(MHz) & $\Delta_f$~(MHz) & repetition\\
\hline
2a & ${\bf P_4}$  &  7.0 & -3.0 & +3.0 & 678 \\
2b& ${\bf C_4}$   &  7.0 & -3.0 & +3.0 & 672 \\
2c& ${\bf C_6}$  &  7.0 & -3.0 & +3.0 & 734  \\
2g& ${\bf S_4}$ &  7.5 & -3.0 & +3.0 & 425 \\
2h& {\bf 3-pan} &  7.5 & -3.0 & +3.0 & 525 \\
2i& {\bf 5-pan} &  7.5 & -3.0 & +3.0 & 570 \\
3a& ${\bf K_5}$ &  7.0 & -3.0 & +1.5 & 1961 \\
3b& ${\bf K_{3,3}}$ &  8.0 & -3.0 & +3.0 & 1279 \\
4c& ${\bf S_6}$ &  7.0 & -3.0 & +3.0 & 1640 \\
\end{tabular}
\end{ruledtabular}
\label{tableS2}
\end{table}

\begin{table}
\caption{Experimental time budget}
\centering
\begin{tabular}{c c }
\hline \hline
Process & Time budget \\
\hline
Single atom loading (PGC) & 100~ms  \\
Initial occupancy checking & $40\times2\times N_p$~ms \\ 
& (for 3D array with $N_p$ planes) \\
Arom rearrangements & 800~ms \\
Final occupancy checking & $40\times2\times N_p$~ms \\
Bias field ON, optical pumping & 20~ms \\
Quantum annealing & $5~\mu$s \\
 & (including trap off) \\
Bias field OFF & 20~ms \\
Final state detection & $40\times2\times N_p$~ms\\
\hline
\end{tabular}
\label{tableS3}
\end{table}

\begin{table*}
\caption{Experimental errors }
\centering
\begin{tabular}{c c }
\hline \hline
Error sources & Treatment \\
\hline
Spontaneous decay from intermediate level $\ket{m}$ & Lindblad master equation \cite{Leseleuc2018,Lee2019} \\
 & $\frac{d\rho}{dt} = -\frac{i}{\hbar}\left [H, \rho  \right ] + \sum_{j=1}^{N}{\left (L_j \rho L_j^{\dagger} - \frac{1}{2}\left \{L_j^{\dagger}L_j, \rho  \right \}  \right )}$ \\
&with $L_j=\sqrt{\gamma_m /2} \sigma_z^{(j)}$ and $\gamma_m = 2\pi\times10$~kHz \\
\hline
Laser phase noise & Monte-Carlo simulation with spectral noises~\cite{Lee2019} \\
 & $10^4$~(Hz${}^2$$/$Hz) (780~nm) and $10^3$~(Hz${}^2$$/$Hz) (480~nm) \\ 
\hline
Laser intensity fluctuation & Monte-Carlo simulation with measured value (2$\%$) \\
\hline
Finite temperature & Monte-Carlo simulation with positional fluctuation~\cite{Jo2020} \\
 & standard deviation $\sigma_r = 0.1~\mu$m (radial) and $\sigma_z = 0.6~\mu$m (axial) \\
& where the temperature 30~$\mu$K, trap depth 1~mK \\
\hline
State-preparation-and-measurement (SPAM) error & Correction of measured microstate probabilities~\cite{Song2021} \\
& $P(\ket{0}|\ket{1}) = 0.18$ and $P(\ket{1}|\ket{0}) = 0.03$ \\
\hline
\end{tabular}
\label{tableS4}
\end{table*}

The experimental time budget is summarized in Table~\ref{tableS3}. After the procedure, resulting atom states are detected with the fluorescence of ground-state atoms during 40 ms cyclic transitions to $\ket{5P_{3/2}, F=3}$, and the procedure is repeated until the probability distribution is obtained. In our experiment, state-preparation-and-detection (SPAM) errors are $P(\ket{0}|\ket{1}) = 0.18$ and $P(1|\ket{0}) = 0.03$~\cite{Song2021}. With these parameters the measured microstates $S$ can be reconstructed to the error-corrected microstates $S'$ by $S'=(M^{-1})^{\otimes N} S$ for $N$ atoms, where $M$ is the SPAM error matrices, given by
\begin{equation}
M = \begin{bmatrix}
1-P(\ket{0}|\ket{1}) & P(\ket{1}|\ket{0})\\ 
P(\ket{0}|\ket{1}) & 1-P(\ket{1}|\ket{0})
\end{bmatrix}.
\label{spam}
\end{equation}

Experimental errors are listed in Table~\ref{tableS4}. Dominant errors are from the spontaneous decay rate of the intermediate state and laser phase noises~\cite{Leseleuc2018,Lee2019}. While the the spontaneous decay from the Rydberg state (of the lifetime $10^4$ times longer than $\ket{m}$) is ignorable, the scattering rate to $\ket{m}$ is $\gamma_m = 2\pi\times10$~kHz. The phase noise of Rydberg excitation lasers are measured by the frequency noise spectrum of the error signal while both lasers are stabilized. The overall spectral densities of frequency noise are measured to $10^4$~(Hz${}^2$$/$Hz) for 780 nm and $10^3$~(Hz${}^2$$/$Hz) for 480 nm around the Fourier frequency near the Rabi frequency $\Omega_0$. Numerical simulations with Lindblad master equation with these effects taken into account are shown (with red dots and purple dots) in Figs.~\ref{fig2}(f,l). Additional experimental errors are from the finite temperature of atoms and the intensity fluctuations of Rydberg excitation lasers. The finite temperature of atoms in the tweezer induces the uncertainty of exact positions of atomic sites. For the temperature 30~$\mu$K and the trap depth 1~mK, the standard deviation of the position is calculated to $\sigma_r = 0.1~\mu$m for radial direction and $\sigma_z = 0.6~\mu$m for axial direction (tweezer propagation direction). The intensity fluctuations of lasers are measured about 2$\%$. Monte Carlo numerical simulation are performed with these effect taken into to account for the probabilities (green dots) in Figs.~\ref{fig2}(f,l).
\end{appendix}

\begin{acknowledgements} \noindent
This research was supported by Samsung Science and Technology Foundation (SSTF-BA1301-52) and National Research Foundation of Korea (NRF) (2017R1E1A1A01074307, 2019M3E4A1080411, 2020R1A4A3079707, and 2021R1A2C4001847).
\end{acknowledgements}


\begin{thebibliography}{99}

\bibitem{Ebadi2021} S. Ebadi, T. T. Wang, H. Levine, H. et al., ``Quantum phases of matter on a 256-atom programmable quantum simulator,'' Nature {\bf 595}, 227 (2021). 

\bibitem{Arute2019} F. Arute, K. Arya, R. Babbush, et al., ``Quantum supremacy using a programmable superconducting processor,'' Nature {\bf 574}, 505 (2019).

\bibitem{Monroe2021} C. Monroe, W. C. Campbell, L.-M. Duan, Z. -X. Gong, A. V. Gorshkov, P. Hess, R. Islam, K. Kim, G. Pagano, P. Richerme, C. Senko, and N. Y. Yao, ``Programmable Quantum Simulations of Spin Systems with Trapped Ions,''  Rev. Mod. Phys. {\bf 93}, 025001 (2021).

\bibitem{Scholl2020} P. Scholl, M. Schuler, H. J. Williams, A. A. Eberharter, D. Barredo, K.-N. Schymik, V. Lienhard, L.-P. Henry, T. C. Lang, T. Lahaye, A. M. L\"{a}uchli, and A. Browaeys, ``Quantum simulation of 2D antiferromagnets with hundreds of Rydberg atoms,'' Nature {\bf 595}, 233 (2021).

\bibitem{Chen2011} H. Chen, X. Kong, B. Chong, G. Qin, X. Zhou, X. Peng, and J. Du, ``Experimental demonstration of a quantum annealing algorithm for the traveling salesman problem in a nuclear-magnetic-resonance quantum simulator,'' Phys. Rev. A {\bf 83}, 032314 (2011).

\bibitem{Yarkoni2018} S. Yarkoni, A. Plaat, and T. Back, ``First results solving arbitrarily structured maximum independent set problems using quantum annealing,” IEEE Congress on Evolutionary Computation (CEC), 1 (2018).

\bibitem{Centrone2021} F. Centrone, N. Kumar, E. Diamanti, et al., ``Experimental demonstration of quantum advantage for NP verification with limited information,'' Nat. Commun. {\bf 12}, 850 (2021). 

\bibitem{Lucas2014}  A. Lucas, ``Ising formulations of many NP problems,'' Frontiers in Phys. {\bf 2}, 5 (2014).

\bibitem{Korte2017} B. Korte and J. Vygen, \emph{Combinatorial Optimization} (Springer, New York, 2017).

\bibitem{Barahona1982} F. Barahona, ``On the computational complexity of Ising spin glass models,'' J. Phys. A: Mathematical and General {\bf 15}, 3241 (1982).

\bibitem{Arora2009} S. Arora and B. Barak, \emph{Computational Complexity: A Modern Approach} (Cambridge University Press, New York, 2009).

\bibitem{Farhi2001} E. Farhi, J. Goldstone, S. Gutmann, J. Lapan, A. Lundgren, and D. Preda, ``A Quantum Adiabatic Evolution Algorithm Applied to Random Instances of an NP-Complete Problem,'' Science  {\bf 292}, 472 (2001).

\bibitem{Dickson2011} N. G. Dickson and M. H. S. Amin, ``Does adiabatic quantum optimization fail for NP-complete problems?'' Phys. Rev. Lett. {\bf 106}, 050502 (2011).

\bibitem{Pichler2018} H. Pichler, S.-T. Wang, L. Zhou, S. Choi, and M. D. Lukin, ``Quantum optimization for maximum independent set using Rydberg atom arrays,'' arXiv:1808.10816 (2018).

\bibitem{Bernien2017} H. Bernien, S. Schwartz, A. Keesling, H. Levine, A. Omran, H. Pichler, S. Choi, A. S. Zibrov, M. Endres, M. Greiner, V. Vuleti\'{c}, and M. D. Lukin, ``Probing many-body dynamics on a 51-atom quantum simulator,'' Nature {\bf 551}, 579 (2017).

\bibitem{Labuhn2016} H. Labuhn, D. Barredo, S. Ravets, S. L\'{e}s\'{e}leuc, T. Macr\'{i}, T. Lahaye, and A. Browaeys, "Tunable two-dimensional arrays of single Rydberg atoms for realizing quantum Ising models," Nature {\bf 534}, 667 (2016).

\bibitem{ISGCI} https://www.graphclasses.org

\bibitem{Kuratowski1930} K. Kuratowski,  ``Sur le probl\'{e}me des courbes gauches en topologie,''  Fund. Math. (in French) {\bf 15}, 271 (1930).

\bibitem{Qiu2020} X. Qiu, P. Zoller, and X. Li, ``Programmable quantum annealing architectures with Ising quantum wires,'' PRX Quantum {\bf 1}, 020311 (2020).

\bibitem{Song2021} Y. Song, M. Kim, H. Hwang, W. Lee, and J. Ahn, ``Quantum simulation of Cayley-tree Ising Hamiltonians with three-dimensional Rydberg atoms," Phys. Rev. Research {\bf 3}, 013286 (2021).

\bibitem{Kim2020} M. Kim, Y. Song, J. Kim, and J. Ahn, ``Quantum-Ising Hamiltonian programming in trio, quartet, and sextet qubit systems,'' PRX Quantum {\bf 1}, 020323 (2020).

\bibitem{Barredo2018} D. Barredo, V. Lienhard, S. L\'{e}s\'{e}leuc, T. Lahaye, and A. Browaeys, "Synthetic three-dimensional atomic structures assembled atom by atom," Nature {\bf 561}, 79 (2018).

\bibitem{Lee2016} W. Lee, H. Kim, and J. Ahn, "Three-dimensional rearrangement of single atoms using actively controlled optical microtraps," Optics Express {\bf 24}, 9816 (2016).

\bibitem{Garey1977} M. Garey and D. Johnson, ``The rectilinear Steiner tree problem is NP-complete,'' SIAM J. Appl. Math. {\bf 32}, 826 (1977).

\bibitem{Preskill2018} J. Preskill, ``Quantum Computing in the NISQ era and beyond,'' Quantum {\bf 2}, 79 (2018).

\bibitem{Denchev2016} V. S. Denchev, S. Boixo, S. V. Isakov, N. Ding, R. Babbush, V. Smelyanskiy, J. Martinis, and H. Neven, ``What is the computational value of finite-range tunneling?'' Phys. Rev. X {\bf 6}, 031015 (2016).


\bibitem{Lee2017} W. Lee, H. Kim, and J. Ahn, ``Defect-free atomic array formation using the Hungarian matching algorithm,'' Phys. Rev. A {\bf{95}}, 053424 (2017).
\bibitem{Kim2019} H. Kim, M. Kim, W. Lee, and J. Ahn, ``Gerchberg-Saxton algorithm for tweezer-trap atom arrangements,'' Opt. Express {\bf{27}}(3), 2184 (2019).


\bibitem{Leseleuc2018} S. de L\'{e}s\'{e}leuc, D. Barredo, V. Lienhard, A. Browaeys, and T. Lahaye, ``Analysis of imperfections in the coherent optical excitation of single atoms to Rydberg states,'' Phys. Rev. A {\bf{97}}, 053803 (2018).

\bibitem{Lee2019} W. Lee, M. Kim, H. Jo, Y. Song, and J. Ahn, ``Coherent and dissipative dynamics of entangled few-body systems of Rydberg atoms,'' Phys. Rev. A {\bf{99}}, 043404 (2019).

\bibitem{Jo2020} H. Jo, Y. Song, M. Kim, and J. Ahn, ``Rydberg Atom Entanglements in the Weak Coupling Regime,'' Phys. Rev. Lett. {\bf 124}, 033603 (2020).

\end{thebibliography}
\end{document}